\documentclass{article}
\usepackage{spconf,amsmath,graphicx,multirow}

\usepackage{amsmath,graphicx}
\usepackage{amssymb,amsmath,epsfig}
\usepackage{lipsum,color,multirow,url}
\usepackage{arydshln, algorithm, algpseudocode}
\usepackage{microtype}
\usepackage{nth}
\usepackage[inline]{enumitem}
\usepackage{romannum}
\usepackage{xcolor}

\title{Benchmarking LF-MMI, CTC and RNN-T criteria for streaming ASR}
\name{
\begin{tabular}{c}
Xiaohui Zhang$^{\star}$, Frank Zhang$^{\star}$, Chunxi Liu$^{\star}$, Kjell Schubert, Julian Chan, Pradyot Prakash, \\
Jun Liu, Ching-Feng Yeh, Fuchun Peng, Yatharth Saraf, Geoffrey Zweig
\end{tabular}
\thanks{$\star$ Equal contribution.}  }
\address{Facebook AI, USA}

\begin{document}
\ninept
\maketitle
\begin{abstract}
In this work, to measure the accuracy and efficiency for a latency-controlled streaming automatic speech recognition (ASR) application,  we perform comprehensive evaluations on three popular training criteria: LF-MMI, CTC and RNN-T. In transcribing social media videos of 7 languages with training data 3K - 14K hours, we conduct large-scale controlled experimentation across each criterion using identical datasets and encoder model architecture. We find that RNN-T has consistent wins in ASR accuracy, while CTC models excel at inference efficiency. 
Moreover,  we selectively examine various modeling strategies for different training criteria, including modeling units, encoder architectures, pre-training, etc. Given such large-scale real-world streaming ASR application, to our best knowledge, we present the first comprehensive benchmark on these three widely used training criteria  across a great many languages.
\end{abstract}

\section{Introduction}
\label{sec:intro}
Thus far there has been a growing interest in speech community to investigate ASR modeling techniques that allow for flat-start (alignment-free) training, 
e.g., connectionist temporal classification (CTC) \cite{graves2006connectionist}, recurrent neural network transducer (RNN-T) \cite{graves2012sequence}, and attention-based sequence-to-sequence (seq2seq) models \cite{chorowski2015attention, chan2016listen}. 
Specifically, CTC criterion learns an encoder-only model, which can be further composed with an $n$-gram language model (LM) in a standard WFST framework. %
RNN-T or attention-based seq2seq model jointly learns an encoder with a neural decoder model that can be considered as a neural LM. 

There have been extensive ASR benchmarks on the public dataset like LibriSpeech \cite{panayotov2015librispeech}. Recently deep transformer based hybrid models have achieved state-of-the-art results among cross-entropy (CE) \cite{wang2019transformerbased} and CTC based models \cite{zhang2020fast} respectively. 
Improving accessibility to social media videos remains an important task, which allows for various applications like automatic video captioning, indexing and retrieval. Transcribing the heterogeneous social media videos of extensively diverse languages is still highly challenging, and prior works \cite{liao2013large, soltau2016neural, chiu2019comparison, liu2020multilingual, zhang2020fast, liu2020contextualizing} have examined various ASR technologies in both real-world high- and low-resource scenarios. 
In this work, we particularly focus on \emph{streaming}  and \emph{long-form}  ASR solutions, where each test utterance is up to 45 seconds long. 
While attention-based seq2seq models have shown difficulty in generalizing well on long utterances given the previous study  \cite{chiu2019comparison},  we thus do not include attention-based seq2seq model in this study.  

To understand the performance distinction of different model types, it is essential to examine whether the model architecture or the training criterion is accountable. This is typically a challenging task, since the model architectures for differing training criteria are intrinsically different, which makes the training criteria not directly comparable. 
However, we note that, for all these training criteria, the majority of model parameters resides in the encoder part. 
In order to understand the relative performance difference in a relatively fair manner,   
we can fix the encoder model architecture for each training criterion to exclude the respective encoder impact on ASR performance. Therefore, in this work, we focus on comparing RNN-T and CTC criteria against a strong hybrid LF-MMI \cite{povey2016purely} baseline, and all cases use the same streamable encoder architecture. We evaluated the modeling performance by word error rate (WER) and decoding efficiency by real-time factor (RTF) of models trained by all three training criteria on video ASR tasks of 7 languages with training data ranging from $3K$ to $14K$ hours, with test sets of three noisy levels for each language. To the best of our knowledge, there has not been such comprehensive study thus far. 
For example, prior work \cite{battenberg2017exploring} provided comparisons among CTC, RNN-T, and attention-based seq2seq models, while not including the standard hybrid ASR model.
Recent work in \cite{chiu2019comparison, li2020comparison} has compared RNN-T with attention-based seq2seq models with the same encoder architecture; however, they used different encoder architectures in comparison to hybrid and CTC models. Besides, \cite{jain2019rnnt} has compared RNN-T with LF-MMI with the same encoder architecture on English.

Additionally, as we understand that RNN-T relies on a neural decoder - which can be seen as an implicit neural LM  - in the first-pass decoding, however, 
all model types are able to explicitly perform additional online neural LM rescoring in the first-pass decoding (e.g. shallow fusion \cite{kannan2018analysis}). So we further exclude any additional LM rescoring effect in this study. 

Besides the comprehensive benchmark of three training criteria, we make further contributions by selectively specifying how we reach the best modeling strategies for each criterion. We present studies on: (i) the choice of modeling units, wordpiece v.s. chenone, and (ii) encoder model architectures, latency-controlled bi-directitonal LSTM (LC-BLSTM) v.s. time-depth separable (TDS) convolutions for CTC models of three languages. We also present optimization efforts - reducing memory consumption and model pre-training - for RNN-T training, which lead to improvements in both modeling performance and training efficiency.

\section{Model training}
For each language we evaluated on, all models were trained on the same data segmented to up-to 10s, which was achieved by force aligning the whole audio against the reference using the same cross-entropy (CE)-trained model. Segmenting training data could substantially improve the training throughput, and slightly improve the accuracy  as shown in \cite{wang2019transformerbased}. LF-MMI models were pre-trained with CE criterion on 10s segments, and then fine-tuned with LF-MMI criterion on 1.5s segments. CTC models were trained on 10s segments directly. The encoder in RNN-T models were pre-trained with CE criterion \cite{li2020comparison} and then the whole model was fine-tuned with RNN-T criterion, all on 10s segments. The chunk size used during training is 1.28s for CTC and RNN-T, and 1.5s for LF-MMI. The right context in training is 210ms for LF-MMI and 240ms for CTC and RNN-T\footnote{The difference in right context is due to it needs to be divisible by the stride e.g. 3, 4 or 8. Empirically the slight different in right context did not affect final WER  and real time factor.}. 

Here we briefly review the three training criteria which are studied in this paper. ASR can be formulated as a sequence-to-sequence problem. Each speech utterance is parameterized as an input acoustic feature vector sequence 
$\textbf{x}  = \{\textbf{x}_1 \ldots \textbf{x}_T\} = \textbf{x}_{1:T} $,
where $\textbf{x}_t \in \mathbb{R}^{d}$ and $T$ is the number of frames in $\textbf{x}$. 
We define a grapheme set or a wordpiece inventory as $\mathcal{Y}$, and the corresponding target sequence of length $U$ as 
$\textbf{y} = \{ y_1 \ldots y_U \} = \textbf{y}_{1:U} $, 
where $y_u \in \mathcal{Y}$. Besides, we define $\bar{\mathcal{Y}}$ as $ \mathcal{Y} \cup \{ \emptyset \}$, where $\emptyset$ is the blank label, and $\bar{\mathcal{Y}}^{*}$ as the set of all sequences over output space $\bar{\mathcal{Y}}$.

\subsection{LF-MMI}
The MMI objective can be formulated as:
\begin{equation}
    F_{MMI} = log \frac{p(\textbf{x}|\textbf{y})}{\sum_{\hat{\textbf{y}}} p(\textbf{x} | \hat{\textbf{y}})}  
    \approx log \frac{p(\textbf{x}|\mathbb{G}_{num})}{p(\textbf{x} | \mathbb{G}_{den})}
\end{equation}
where $\hat{\textbf{y}}$ represents any possible hypothesis. In LF-MMI \cite{povey2016purely}, a composite HMM graph called ``denominator graph" $\mathbb{G}_{den}$ is used to approximate the denominator, which encodes all possible hypothesis, and thereby we have $\sum_{\hat{\textbf{y}}} p(\textbf{x} | \hat{\textbf{y}}) \approx  p(\textbf{x} | \mathbb{G}_{den})$. Efficient computation of the denominator without having to generate lattices is enabled by adopting an n-gram phone/character language model (LM) when generating $\mathbb{G}_{den}$, and doing full forward-backward computation on GPUs. Similarly, the numerator $p(\textbf{x}|\textbf{y})$ is approximated by $p(\textbf{x}|\mathbb{G}_{num})$ where $\mathbb{G}_{num}$ is another composite HMM graph called ``numerator graph", encoding all possible sequences of HMM states pertaining to the transcription $\textbf{y}$. It could be either an acyclic graph encoding pre-computed alignments, giving regular LF-MMI (used in this work), or a graph with self-loops determined solely by reference transcripts, giving flat-start (alignment-free) LF-MMI.

\subsection{CTC}
As a sequence-level training criterion, for CTC, the log-likelihood of a given target sequence $\textbf{y}$ can then be found by summing the probabilities of all allowed alignments. Specifically, 
\begin{align}
\label{ctc_formula}
    \log p(\textbf{y}|\textbf{x}_{1:T}) = \sum_{\textbf{a} \in \mathcal{B}^{-1}(\textbf{y}) }\prod_{t=1}^{t=T}{p(\textbf{a}_t|\textbf{x}_t)}
\end{align}
where $\mathcal{B}: \bar{\mathcal{Y}}^* \rightarrow  \mathcal{Y}^{*}  $ is a mapping operation that removes all blank labels and repeating symbols in a given sequence. The encoder is trained to maximize the log-likelihood for each training example and $p$ can be computed efficiently using the forward-backward algorithm.

Note that the underlying assumption in Eq (\ref{ctc_formula}) is that probabilities between timestamps are conditional independent. The Transducer criterion introduced in the next section will lift this constraint. 
\subsection{RNN-T}
\label{ssec:rnn-t}


Excluding the conditional independence assumption made in CTC, RNN-T models the posterior probability as:
\begin{equation}
    P( \textbf{y}  | \textbf{x}) =  \sum\limits_{ \textbf{a}  \in \mathcal{B}^{-1}(\textbf{y} ) }    P( \textbf{a}  | \textbf{x})
\end{equation}
\noindent 
where $\mathcal{B}: \bar{\mathcal{Y}}^* \rightarrow  \mathcal{Y}^{*}  $ is a mapping operation that removes all blank labels in a given sequence. RNN-T model parameterizes the alignment probability $P(\textbf{a} | \textbf{x})$ and computes it with an encoder network (i.e. transcription network in \cite{graves2012sequence}), a prediction network and a joint network. 
The encoder performs a mapping operation, denoted as $f^{\text{enc}}$,  which converts $\textbf{x}$ into another sequence of representations $\textbf{h}^{\text{enc}} = \{ \textbf{h}_1^{\text{enc}} \ldots \textbf{h}^{\text{enc}}_{T'} \}$: 
\begin{equation}
 \textbf{h}^{\text{enc}} = f^{\text{enc}}(\textbf{x}) 
\end{equation}
\noindent where $T'$ is equal or shorter than $T$ due to subsampled frame rate. 
A  prediction network $f^{\text{pred}}$, based on RNN or its variants, takes both its state vector and the previous non-blank output label $y_{u-1}$ predicted by the model, to produce the new representation $\textbf{h}^{\text{pred}}$: 
\begin{equation}
    \textbf{h}^{\text{pred}}_{1:u} = f^{\text{pred}}(y_{0:(u-1)})
\end{equation}
\noindent where $u$ is the output label index and $y_0 = \emptyset$.
Finally, the joint network $f^{\text{join}}$  is a feed-forward network that combines the encoder output $\textbf{h}^{\text{enc}}_t$ and prediction network output $\textbf{h}^{\text{pred}}_u$ to compute logits $\textbf{z}_{t,u}$, which go through a softmax function and produce a posterior distribution of the next output label over $\bar{\mathcal{Y}}$:
\begin{equation}
    \textbf{z}_{t,u} = f^{\text{join}}(\textbf{h}^{\text{enc}}_t, \textbf{h}^{\text{pre}}_u) 
    \label{eq:joiner}
\end{equation}
\begin{equation}
    p(y_u| \textbf{x}_{1:t}, y_{1:(u-1)}) = \text{Softmax}(\textbf{z}_{t,u})
\label{eq:posterior}
\end{equation}
\noindent 
The encoder can be seen as an acoustic model, and the combination of prediction and joint network as a decoder. 


\section{Modeling units}
For the LF-MMI criterion, we used tied context-dependent grapheme states (i.e. chenones) \cite{le2019senones} with a stride (sub-sampling factor) of 3. For CTC criterion, we used wordpiece units with a stride of 8. For RNN-T criterion, we used wordpiece units with a stride of 4. For each training criterion, the choice of modeling units, and stride were tuned separately on validation data to achieve the best balance between WER and inference efficiency. In Section \ref{wp-vs-chenone}, we will specify our analysis in the modeling unit options for CTC. Besides, the size of the chenone set (i.e. decision tree size) or wordpiece vocabulary were tuned to optimize WER on each language for each model, also on validation data.

\section{Model architecture}
We keep the encoder architecture fixed when comparing performance across different training criteria. In the main experiment, we used a latency-controlled bi-directitonal LSTM (LC-BLSTM) encoder with 5 layers of 800 hidden units. Sub-sampling along the time dimension by a factor of 3 is applied at the output of the first layer to achieve a stride of 3 for LF-MMI models, and sub-sampling by a factor of 2 is applied at the output of first, second or third layer to achieve a stride of 4 or 8 respectively for RNN-T and CTC models. The encoder alone has around 75M parameters. 
All the models presented here can run in a streaming fashion, because of the limited right context.

\section{Model inference}
For LF-MMI and chenone-CTC models, we pre-built decoding graphs $\rm{H \circ C \circ L}$ and $\rm{G}$ \footnote{H transduces HMM states to context-dependent graphemes; C transduces context-dependent graphemes to graphemes; L transduces graphemes to words; G represents the language model.} and dynamically composed them during decoding \cite{liu2019efficient}. For wordpiece-CTC models, we use the same dynamic decoding approach but pre-built $\rm{H \circ L}$ rather than $\rm{H \circ C \circ L}$ \cite{zhang2020fast}. For RNN-T model, we use standard beam search decoding without any external LM fusion as in \cite{graves2012sequence}, since we are evaluating the three training criteria under a ``vanilla" single-pass inference setting without any LM fusion/rescoring for all models. All acoustic models were trained in PyTorch and applied post-training INT8 quantization to enable efficient decoding. Decoding hyper-parameters, e.g. beam sizes, were tuned on validation data for each model separately, to achieve a balance between decoding efficiency and WER. To satisfy latency constraints for live captioning use-case, we limit the chunk size to 0.8s for English, Spanish, Hindi and Indic English, and 1.5s for Thai, Vietnamese and Turkish across all models. 
\section{Experiments}
\subsection{Data}
We evaluate all models on our in-house Video ASR datasets, which are sampled from public social media videos and completely de-identified before transcription; both transcribers and researchers do not have access to any user-identifiable information (UII). These videos contain a diverse range of speakers, accents, topics, and acoustic conditions making automatic recognition very challenging. We included a wide variety of languages in this study in order to get a broad understanding of model performance, including: (i) fusional languages Spanish (ES), Hindi (HI) and Indic English (EN-IN), (ii) analytic languages US English (EN-US), Vietnamese (VN) and Thai (TH), and (iii) an agglutinative language Turkish (TR). The training set sizes are shown in Table \ref{duration}. Note that we combined Hindi and Indic English training data and trained a single model for each criterion, although we evaluate the model on Hindi and Indic English test sets separately. The reason is that due to their similarity in pronunciation and frequent code-switching, it can be hard for a language identification (LID) model to differentiate acoustic inputs from these languages. In addition, some special text processing was applied to Thai: as reference transcripts were un-segmented (no word-level tokenization), we needed to tokenize the transcripts at wordpiece level by training a wordpiece model first, and then construct a lexicon mapping wordpieces to graphemes for data segmentation, model training and decoding. Regarding data augmentation, speed perturbation \cite{ko2015audio} and \emph{SpecAugment} \cite{park2019specaugment} (LD policy for RNN-T and SM policy for CTC and LF-MMI) are used. 

The test sets for each language are composed of \texttt{clean}, \texttt{noisy} and \texttt{extreme} categories, with \texttt{extreme} being the most acoustically challenging. The validation set for each language was composed of data from the \texttt{noisy} category. The duration of validation and test sets for each category in each language is around 10 to 40 hours. 
All validation and test data were segmented up to 45s. 

\begin{table}[htb]
    \centering
  \small
      \caption{Training data sizes (in hours). }
    \begin{tabular}{*7c}
    \hline
    EN-US & ES & HI $\&$ EN-IN & TH & VN & TR \\
    \hline\hline
    14K & 7.2K & 6.7K & 5.1K & 4.2K & 3.1K  \\
    \hline
    \end{tabular}
    \label{duration}
\end{table}
\vspace{-1em}

\subsection{Results}
In this section, we first present decoding results on all 7 languages in Table \ref{overview}, with the best overall modeling strategies (modeling unit, stride, and pre-training strategy) chosen for each training criterion. For each language, all models were trained on the same data with the same encoder model architecture (5$\times$800 LC-BLSTM). Later, we will selectively analyze the impact of modeling strategies and encoder model architecture for CTC/RNN-T.
We use word error rate (WER), or character error rate (CER) for Thai, to measure modeling performance on \texttt{clean}, \texttt{noisy} and \texttt{extreme} test splits for each language and model. We use real-time factor (RTF) to measure decoding efficiency.  

\label{sec:format}
\setlength{\tabcolsep}{0.14cm}
\begin{table*}[t]
\caption{ {\it Performance overview of WER (CER for Thai) and RTF. Average WERR (Word Error Rate Reduction, positive and larger is better) is computed by first computing the WERR on the three test categories individually (using LF-MMI models as a baseline), and then taking the unweighted average.}}
\begin{tabular}{  p{2.2cm} | c c c | c c c | c c c | c c c }
Language & \multicolumn{3}{c}{US English} & \multicolumn{3}{|c|}{Spanish} & \multicolumn{3}{c}{Hindi} & \multicolumn{3}{|c}{Indic English} \\
\cline{1-13}  
Model &  \small{LF-MMI} &	 \small{CTC} &	 \small{RNN-T} 
&  \small{LF-MMI} &	 \small{CTC} &	 \small{RNN-T} 
&  \small{LF-MMI} &	 \small{CTC} &	 \small{RNN-T} 
&  \small{LF-MMI} &	 \small{CTC} &	 \small{RNN-T}  \\
\hline
\hline
\small{\texttt{clean}} & 10.4  & 11.3  & 10.2  & 10.4 & 10.2 & 9.1   & 20.1 & 18.9 & 17.9   & 26.9 & 26.7 & 26.2 \\
\small{\texttt{noisy}} & 14.4  & 15.0  & 14.2   & 12.7 & 12.6 & 11.1   & 21.7 & 20.6 & 19.4   & 31.6 & 31.1 & 31.3 \\
\small{\texttt{extreme}} & 20.3 & 20.9 & 19.8  & 21.0 & 20.7 & 19.2  & 25.7 & 26.3 &  25.0  & 32.2 & 32.7 & 31.3 \\ \cdashline{1-13}[4.0pt/0.5pt]
\small{Avg. WERR}    & -- & -5.3$\%$ & 1.9$\%$  & -- & 1.4$\%$ & 11.2 $\%$  & -- & 2.9$\%$  & 8.1$\%$   & -- & 0.3$\%$ & 2.1$\%$  \\
\hline
\small{RTF} & 0.46 & 0.40 & 0.49  & 0.50 & 0.33 & 0.48   & 0.44 & 0.30 & 0.41   & 0.44 & 0.30 & 0.41 \\
\hline
\end{tabular}
\vspace{0.3cm}
\newline
\centering
\begin{tabular}{  p{2.2cm} | c c c | c c c | c c c} 
Language & \multicolumn{3}{c}{Thai} & \multicolumn{3}{|c|}{Vietnamese} & \multicolumn{3}{c}{Turkish} \\
\cline{1-10}  
Model &  \small{LF-MMI} &	 \small{CTC} &	 \small{RNN-T} 
&  \small{LF-MMI} &	 \small{CTC} &	 \small{RNN-T} 
&  \small{LF-MMI} &	 \small{CTC} &	 \small{RNN-T}  \\
\hline
\hline
\small{\texttt{clean}} &9.7  & 9.9 & 8.7 & 11.5 & 11.7 & 10.5 & 19.4 & 19.6 & 16.9      \\
\small{\texttt{noisy}} & 13.7  &  14.2 & 12.8 & 19.3 & 19.9 & 19.0   & 20.2 & 20.7 & 18.6   \\
\small{\texttt{extreme}} & 21.7  & 22.8 & 20.2  & 45.3 & 46.6 & 46.3   & 37.9 & 39.9 & 38.4  \\
\cdashline{1-10}[4.0pt/0.5pt] 
\small{Avg. WERR}  & -- & -3.6$\%$  & 7.9$\%$   & -- & -2.6$\%$ & 2.6$\%$  & -- & -2.9$\%$ & 6.5$\%$  \\

\hline
\small{RTF} & 0.41 & 0.29 & 0.40  & 0.37 & 0.29 & 0.44   & 0.45 & 0.33 & 0.43    \\
\hline
\end{tabular}
\label{overview}
\vspace{-1em}
\end{table*}
\subsection{WP-CTC v.s. chenone-CTC }
\label{wp-vs-chenone}
It is well-studied that LF-MMI works well with chenone units and RNN-T works well with wordpiece units. For CTC training, we have observed that full sequence deep transformer encoder trained with chenone units outperforms wordpieces consistently. However, the trend is different for our streaming application in this work. In Table \ref{units}\footnote{WERs of CTC in Table \ref{units} and Table \ref{encoder} are different from Table \ref{overview} due to some differences in evaluation datasets, however the same data were used consistently within each table.}, we show results of WP-CTC (with a stride of 8) and chenone-CTC (with a stride of 4) on English, Vietnamese and Turkish, which are the best choices of stride for WP/chenone-CTC respectively, in terms of balancing WER and RTF. We find that for LC-BLSTM models, WP-CTC training consistently outperforms chenone-CTC in WER\footnote{and also in RTF \cite{zhang2020fast}, though we did not measure RTF here.}, and thus we decided to adopt WP-CTC in comparison with other training criteria. One hypothesis is that the LC-BLSTM encoder being a streaming model is less expressive than a full-context deep transformer encoder. Therefore, the LC-BLSTM encoder is not able to fully exploit the richer target representation provided by chenone alignments during training, in a sense that the optimal size of a chenone is usually much larger than a wordpiece set. 

\begin{table}[]
    \centering
    \small
        \caption{WER of WP-CTC v.s. chenone-CTC} 
    \begin{tabular}{  p{1.2cm} | p{0.3cm} c | p{0.3cm} c | p{0.3cm} c}
Lang. & \multicolumn{2}{c}{EN} & \multicolumn{2}{|c|}{VN} & \multicolumn{2}{c}{TR} \\
\cline{1-7}  
Unit &  \small{WP} &	 \small{chenone}
&  \small{WP} &	 \small{chenone}
&  \small{WP} &	 \small{chenone} \\
\hline
\hline
\small{\texttt{clean}} & 14.0 & 15.3 & 11.6 & 15.5 & 19.3 & 20.7 \\

\small{\texttt{noisy}} & 20.0 & 21.3 & 19.9 & 23.5 & 20.4 & 21.5 \\

\small{\texttt{extreme}}  & 26.1 & 28.5 & 46.6 &  52.2 & 39.9 & 40.6 \\
\hline
\end{tabular}
\label{units}
\end{table}

\subsection{Choices of the encoder model architecture}
We also explored an encoder architecture based on time-depth separable (TDS)
convolutions \cite{hannun2019sequence} as an alternative encoder choice under the CTC setup. Since from recent research \cite{pratap2020scaling} TDS encoder has shown its advantage of speed during inference, we want to explore if this architecture generalizes well on more datasets. The TDS architecture in this study is designed to use as many parameters as possible to improve WER given that the RTF is still lower than LC-BLSTM: the TDS encoder consists of 14 TDS blocks, 3 sub-sampling layers, each with a stride of 2, for a total sub-sampling factor of 8. The total right context is 570 ms. Total parameters is 122M  and is larger than the LC-BLSTM model (75M parameters). The realized RTF of TDS on English (0.26 \cite{pratap2020scaling}) is lower than LC-BLSTM as in Table \ref{overview}. Results on English, Vietnamese and Turkish languages are presented in Table \ref{encoder}. We can see that TDS WER slightly outperforms LC-BLSTM in English and lags behind in the other two languages. One possible explanation is that the TDS architecture is more data hungry, e.g. for English, there are more than 3 times the training data as Turkish and Vietnamese. There is also evidence that on LibriSpeech which has 1000hrs of training data, LC-BLSTM is outperforming TDS  \cite{hannun2019sequence, le2019senones}. Therefore, for the overall 7 languages comparisons, we used the LC-BLSTM encoder when comparing different training criteria. 

\setlength{\tabcolsep}{0.17cm}
\begin{table}[]
    \centering
    \small
        \caption{WER of LC-BLSTM v.s. TDS encoder for CTC}
    \begin{tabular}{  p{1.2cm} | p{0.7cm} c | p{0.7cm} c | p{0.7cm} c}
Lang. & \multicolumn{2}{c}{EN} & \multicolumn{2}{|c|}{VN} & \multicolumn{2}{c}{TR} \\
\cline{1-7}  
Unit &  \scriptsize{LC-BLSTM} &	 \scriptsize{TDS}
& \scriptsize{LC-BLSTM} &	 \scriptsize{TDS}
& \scriptsize{LC-BLSTM} &	 \scriptsize{TDS} \\
\hline
\hline
\small{\texttt{clean}} & 14.0 & 13.7 & 11.6 & 12.7 & 19.3 & 20.9 \\

\small{\texttt{noisy}} & 20.0 & 19.5 & 19.9 & 20.9 & 20.4 & 22.4 \\

\small{\texttt{extreme}}  &  26.1 & 25.2 & 46.6 &  48.2 & 39.9 & 41.8 \\
\hline
\end{tabular}
\label{encoder}
\end{table}

\begin{table}[t]
\caption{\label{tab:data} {\it Training optimization and CE pre-training effects for RNN-T with varying training mini-batch size. WER results on Turkish (without model quantization and decoding beam sweeping). }}
\centering 
\vspace{1em}
    \begin{tabular}{  c | c c | c c }

                  batch size  &    \multicolumn{2}{c|}{8}  &   \multicolumn{2}{c}{16}  \\ 
 \hline
     pre-training                &      N      &    Y    &    N      &     Y    \\
 \hline 
 	\small{\texttt{clean}}      & 17.7     &     17.0   &   17.1  &  16.8  \\    
     \small{\texttt{noisy}}      & 19.3     &     19.0   &   18.9  &  18.9 \\  
\hline
\end{tabular}
 \label{pretrain}
\end{table}
\subsection{(Pre-)training optimization for RNN-T }

One of the major challenges of training RNN-T models is the need of enormous memory size, due to the formulation on both embeddings from the encoder $\textbf{h}^{\text{enc}}_t$ and the predictor $\textbf{h}^\text{pre}_u$ as shown in Eq. \ref{eq:joiner}.
Specifically, in order to compute the forward-backward algorithm \cite{graves2012sequence}, a joint embedding $\textbf{z}_{t,u}$ is needed for each position pair $(t, u)$.
This translates to a minimum memory usage of $T_i * U_i * D$ floating numbers for the $i$-th sequence in a sequence of batch size $B$, where $T_i$ and $U_i$ are sequence lengths of encoder/predictor embeddings and $D$ is the number of sentence pieces as output units.
This can in turn lead to $B * \text{max}_i(T_i) * \text{max}_i(U_i) * D$ floating numbers for the entire batch if with the more traditional ``broadcasting" implementation, or the reduced ${\textstyle\sum}_{i} T_i * U_i * D$ with optimization \cite{li2019improving}.  
For either cases, the scale of such tensors is often measured in GBs, therefore limits the batch size, which is observed to be highly correlated with the stability of gradients and then the final word error rates.

With the identification of the bottleneck for training RNN-T models, our in-house RNN-T criterion implementation provides additional improvements on training efficiency and word error rate reduction with highly optimized memory consumption.
First, {\it function merging} \cite{li2019improving} was adopted to fuse the softmax operation into the RNN-T criterion, this reduces the memory usage by $\simeq50\%$ (translating to 2x batch size) while the numerical value of gradients stay identical.
Second, mixed-precision training was implemented in which 16-bit float numbers (fp16) are used instead of 32-bit ones (fp32), which leads to another $\simeq50\%$ memory usage reduction (another 2x gain on batch size), with some loss on precision but compensated later by larger batch sizes.
The combined optimizations improves the batch size by a factor of 4 compared over the vanilla implementation, and a factor of 2 over function merging alone, which leads to not only training speed-up but also performance gain. 

With an output wordpiece size 2048,  RNN-T training can only use a batch size 8 in each V100 GPU of 16G memory before the above training optimization.
In such case,  pretraining the RNN-T encoder with the hybrid CE model (i.e. the same model used in LF-MMI systems) has provided consistent performance improvements, as shown in Table \ref{pretrain}. 
After the training optimization enables a batch size 16, we observe noticeable performance improvements without encoder pre-training, while additional pre-training with hybrid CE model only provides minor further gains.


\section{Discussions and Conclusions}
\label{sec:con}
In this work, we demonstrated in details that across the 7 languages studied, CTC systems achieved best decoding efficiency while RNN-T systems provided best WER overall. Compared with the LF-MMI baseline, for CTC, the RTF improvement is around $30\%$ with $2$ - $5\%$ WER degradation for 4 languages and up to $3\%$ WER improvement for the other 3 languages; for RNN-T, the RTF is about the same as baseline LF-MMI systems for all languages, with significant and consistent $2$ - $11\%$ WER improvements.

Overall, CTC systems were able to achieve the best decoding efficiency since they use wordpiece units (spanning longer temporal space than chenone) with the largest stride 8 among all systems. This agrees with the findings in prior works that CTC can achieve a good trade-off between WER and decoding inference efficiency when using wordpieces \cite{9003834,das2018advancing,zhang2020fast} or even whole words \cite{soltau2016neural} as modeling units. RNN-T systems consistently achieved the best WERs across all languages, presumably due to to its expressiveness in explicitly leveraging previous output labels, as shown in Eq. \ref{eq:posterior}. 

In future work, we will continue to measure the performance on named entities (i.e. entity error rate), and 
present studies on the ASR inference latency, i.e., delayed token generation problem \cite{inaguma2020minimum, mahadeokar2021alignment}. We will also examine various model-specific techniques that can improve a model type in particular \cite{tjandra2020deja,liu2021improving,pandey2021dual}, and continue to benchmark the best systems across training criteria. 

In summary, each system explored in this study - LF-MMI, CTC, and RNN-T - has its own 
strengths and limits, and accordingly each could be adopted based on different business requirements, e.g. prioritizing run time over WER, or vise versa. 

\bibliographystyle{IEEEbib}
\bibliography{strings,refs}

\end{document}